\documentclass[preprint]{elsarticle}

\usepackage{amsmath,graphicx,siunitx,xcolor,hyperref,physics-patch,bm}
\hypersetup{
colorlinks,
linkcolor={blue!50!black},
citecolor={blue!50!black},
urlcolor={blue!80!black}}
\newcommand{\NDOF}{N_\mathrm{DOF}}

\begin{document}
\title{Benchmarking thermostat algorithms in molecular dynamics simulations of a binary Lennard-Jones glass-former model}
\author{Kumpei Shiraishi}
\ead{kumpei.shiraishi@sanken.osaka-u.ac.jp}
\author{Emi Minamitani}
\address{SANKEN, the University of Osaka, Suita, Osaka 567-0047, Japan}
\author{Kang Kim}
\address{Division of Chemical Engineering, Department of Materials Engineering Science, Graduate School of Engineering Science, the University of Osaka, Toyonaka, Osaka 560-8531, Japan}

\begin{abstract}
A systematic comparison was carried out to assess the influence of representative thermostat methods in constant-temperature molecular dynamics simulations.
The thermostat schemes considered include the Nosé--Hoover thermostat and its chain generalisation, the Bussi velocity rescaling method, and several implementations of the Langevin dynamics.
Using a binary Lennard-Jones liquid as a model glass former, we investigated how the sampling of physical observables, such as particle velocities and potential energy, responds to changes in time step across these thermostats.
While the Nosé--Hoover chain and Bussi thermostats provide reliable temperature control, a pronounced time-step dependence was observed in the potential energy.
Amongst the Langevin methods, the Grønbech-Jensen--Farago scheme provided the most consistent sampling of both temperature and potential energy.
Nonetheless, Langevin dynamics typically incurs approximately twice the computational cost due to the overhead of random number generation, and exhibits a systematic decrease in diffusion coefficients with increasing friction.
This study presents a broad comparison of thermostat methods using a binary Lennard-Jones glass-former model, offering practical guidance for the choice of thermostats in classical molecular dynamics simulations.
These findings provide useful insights for diverse applications, including glass transition, phase separation, and nucleation.
\end{abstract}
\maketitle

\section{Introduction}
Molecular dynamics (MD) simulation is a fundamental computational method in physics, chemistry, biology, and engineering for investigating the properties of many-body systems~\cite{Allen_Tildesley,Frenkel_Smit,TuckermanSM}.
Since its origins in the 1950s~\cite{Alder1957}, MD has undergone continuous methodological development.
More recently, progress in data science has accelerated its development, most notably through machine-learning potentials that achieve near \textit{ab initio} accuracy while being orders of magnitude faster than density functional theory~\cite{Behler_2007}.
Nevertheless, the core algorithms used to generate statistical ensembles have remained largely unchanged, and the reliability of numerical results continues to depend critically on the underlying integration scheme.

In its simplest form, MD integrates Newton's equations of motion and generates the microcanonical (\textit{NVE}) ensemble.
In practice, however, most applications require constant-temperature conditions, corresponding to the canonical (\textit{NVT}) ensemble, which has led to the development of various thermostat algorithms.
Early approaches, such as simple velocity rescaling~\cite{WOODCOCK1971257} or the Berendsen thermostat~\cite{Berendsen1984}, are straightforward to implement but fail to reproduce the canonical ensemble.
More rigorous deterministic approaches include the Nosé--Hoover thermostat~\cite{Nose_JCP_1984,Nose_MolPhys_1984,Hoover_1985} and its chain generalisation~\cite{Martyna_1992}, which generate the canonical ensemble through an extended Hamiltonian formalism.

An alternative approach is the stochastic formulation based on the Langevin dynamics~\cite{Schneider1978,BRUNGER1984495}, whose associated Fokker--Planck equation admits the canonical distribution as its stationary solution.
In the Langevin equation, the force on each particle comprises the interaction force as well as friction and stochastic components.
In this context, the Langevin dynamics is often regarded as the local thermostat.
Recent studies have focused on accurate and stable discrete-time integrators for Langevin dynamics~\cite{VANDENEIJNDEN2006310,Leimkuhler2012Rational,GJF2013}.
Among these, operator-splitting methods have proven particularly effective~\cite{Leimkuhler2012Rational,Leimkuhler2013Robust}, analogous to the familiar `kick-drift-kick' scheme of the Verlet integration~\cite{TuckermanSM}.
In this framework, the system's Liouville operator is decomposed into three operators, `A' (`drift', updating positions), `B' (`kick', updating momenta), and `O' (the Ornstein--Uhlenbeck process), with the BAOAB decomposition widely regarded as especially accurate~\cite{Leimkuhler2012Rational,Leimkuhler2013Robust}.
By contrast, the Grønbech-Jensen--Farago (GJF) method~\cite{GJF2013} represents a direct discretisation of the Langevin equation, designed to generate correct configurational sampling and diffusion.
Subsequent studies~\cite{2GJ2019,GJ2020} further demonstrated that, with an appropriate definition of the half-step velocity as in leap-frog integration, the method can also reproduce the correct velocity distribution.
Moreover, a recent study extended the half-step velocity approach to the isothermal-isobaric (\textit{NPT}) ensemble, achieving improved sampling in velocity space~\cite{Jung2025Langevin}.

In the Langevin equation, the system's dynamical properties are strongly influenced by the choice of the friction coefficient since the friction and stochastic terms of the Langevin equation substantially affect the underlying Hamiltonian dynamics.
Bussi and coworkers proposed a stochastic velocity-rescaling method (hereafter referred to as Bussi thermostat), extending the Berendsen thermostat by incorporating a stochastic term controlling the total kinetic energy, designed to minimise the disturbance on the Hamiltonian dynamics~\cite{Bussi_2007}.
This velocity-rescaling method has been shown to correspond to the global thermostat form of the Langevin dynamics~\cite{BUSSI200826}.
The effectiveness of the Bussi thermostat has been demonstrated in MD simulations of TIP4P liquid water and Lennard-Jones fluids~\cite{Bussi_2007,BUSSI200826}.
Furthermore, the properties of the Bussi thermostat have been systematically compared with those of the velocity scaling method, Nosé--Hoover thermostat, and Langevin dynamics~\cite{Braun2018}.

Despite the development of numerous thermostat algorithms, systematic evaluations of their sampling performance remain limited.
In particular, as noted above, although several discretisation schemes for Langevin dynamics have been proposed, comprehensive assessments of their accuracy and efficiency are still scarce.
A balanced evaluation across algorithms therefore provide practical guidance for the thermostat selection in MD simulations.
In this study, we focus on a realistic yet conceptually simple model system: the Kob--Andersen binary Lennard-Jones mixture~\cite{Kob_Andersen_I_1995}.
Originally developed to mimic a nickel-phosphor metallic glass, it is now widely regarded as the standard model for studying the glass transition and supercooled liquids~\cite{Pedersen_2018}.
Using this system as a benchmark, we systematically compare seven representative thermostat schemes, namely, Nosé--Hoover chains (with one and two degrees of freedom), the Bussi thermostat, and four variants of Langevin thermostats, under identical initial conditions and simulation settings.
We evaluate a set of statistical observables, including temperatures, potential energies, a structural quantity, and dynamical relaxations.
Notably, because the dynamics of glass-forming liquids slow down with decreasing temperature while their static structures remain similar to those of the normal liquids~\cite{Debenedetti_2001}, thermostat effects on static and dynamical properties constitute an important aspect of MD simulations in glass-forming systems~\cite{Gleim_1998,Szamel2004EPL,Berthier_JCP_I_2007,Berthier_JCP_II_2007,Coslovich_EPJE_2018}.

The rest of the paper is structured as follows.
In Sec.~\ref{sec:methods}, we introduce the model system and describe the simulation protocol, with primary focus on particle velocities and potential energies.
In Sec.~\ref{sec:temperature}, we examine the probability distribution of temperatures in the sampled configurations and the associated velocities, which are expected to follow the Maxwell--Boltzmann distribution.
Section~\ref{sec:potential} focuses on the potential energy, while Sec.~\ref{sec:rdf} addresses structural properties, with particular attention to errors induced by the integration time step.
In Sec.~\ref{sec:dynamics}, we analyse the relaxation dynamics under each thermostat scheme.
Section.~\ref{sec:lowT} presents a validation of our main results at a low temperature $T = 0.5$.
Finally, in Sec.~\ref{sec:conclusion}, we discuss the practical implications of our findings, including the computational cost associated with each algorithm.

\section{Numerical methods}
\label{sec:methods}
\subsection{Model}
We study binary Lennard-Jones particles throughout this article.
Particles $i$ and $j$ interact with the Lennard-Jones potential
\begin{align}
 u(r_{ij}) = 4\epsilon_{ij} \bqty{\pqty{\frac{\sigma_{ij}}{r_{ij}}}^{12} - \pqty{\frac{\sigma_{ij}}{r_{ij}}}^6},
\end{align}
where $r_{ij}$ is the distance between the two particles.
In the system, there are two types of particles (denoted as A and B) with a ratio of 80:20.
Particles A and B have an identical mass $m$.
We consider the three-dimensional system ($d = 3$) employing periodic boundary conditions for all directions.
The linear size of the simulation box $L$ is determined by the number density $\rho = N/L^d = 1.2$.
All simulations are performed with $N = 1000$ ($N_\mathrm{A} = 800$ and $N_\mathrm{B} = 200$).
We set the parameters $\epsilon_{ij}$ and $\sigma_{ij}$ with the famous Kob--Andersen mixture~\cite{Kob_Andersen_I_1995}:
$\sigma_\mathrm{AB} / \sigma_\mathrm{AA} = 0.8$, $\sigma_\mathrm{BB} / \sigma_\mathrm{AA} = 0.88$, $\epsilon_\mathrm{AB} / \epsilon_\mathrm{AA} = 1.5$, $\epsilon_\mathrm{BB} / \epsilon_\mathrm{AA} = 0.5$.
Lengths, energies, mass, and times are measured in units of $\sigma_\mathrm{AA}$, $\epsilon_\mathrm{AA}$, $m$, and $\pqty{m\sigma_\mathrm{AA}^2/\epsilon_\mathrm{AA}}^{1/2}$, respectively.
The Boltzmann constant $k_B$ is set to unity.

At the cutoff distance $r_{ij,c}$, we smooth the potential up to its first derivative via
\begin{align}
U = \sum_{i,j} \bqty{
u(r_{ij}) - u(r_{ij,c}) - u^\prime(r_{ij,c}) \pqty{r_{ij} - r_{ij,c}}
},
\end{align}
where $u^\prime(r_{ij,c})$ is the first derivative of the potential at $r_{ij,c}$.
Unlike the original Kob--Andersen model, the cutoff distance is set to $r_{ij,c} = 1.5\sigma_\mathrm{AA}$ for AA and BB interactions and $r_{ij,c} = 2.5\sigma_\mathrm{AB}$ for AB interaction~\cite{Schroder_2020}.
Note that this modification gives identical properties with the original system and yet suppresses computational time thanks to its smaller cutoff radius~\cite{Toxvaerd2011Communication}.

\subsection{Integrators}
In this article, we compare seven integration schemes to generate the canonical (\textit{NVT}) ensemble, namely
\begin{enumerate}
 \item Nosé--Hoover thermostat~\cite{Nose_JCP_1984,Nose_MolPhys_1984,Hoover_1985} (NHC1),
 \item Nosé--Hoover chain thermostat~\cite{Martyna_1992} with the chain size $M = 2$ (NHC2),
 \item Bussi thermostat~\cite{Bussi_2007} (Bussi),
 \item Langevin thermostat with the BAOAB operator splitting~\cite{Leimkuhler2012Rational} (BAOAB),
 \item Langevin thermostat with the ABOBA operator splitting~\cite{Leimkuhler2012Rational} (ABOBA),
 \item Langevin thermostat with the stochastic position Verlet method~\cite{SKEEL20122002,Melchionna2007} (SPV),
 \item Langevin thermostat with the GJF algorithm and the half-step velocity~\cite{Grønbech-Jensen2023} (GJF).
\end{enumerate}
The abbreviations in parentheses are used to represent each scheme.
For NHC1 and NHC2, we implement the time-reversible form of the Nosé--Hoover chain method proposed by Martyna and coworkers~\cite{Martyna_1996}.
Note that the Nosé--Hoover thermostat is equivalent to the Nosé--Hoover chain thermostat with the number of chain $M = 1$.
As an integration scheme, we implement the velocity Verlet algorithm~\cite{Swope1982} for the NHC thermostats as well as the Bussi thermostat.

For the Langevin thermostats, we study four different algorithms.
In addition to the two schemes derived from operator splitting (BAOAB and ABOBA), we also implement the SPV method, which was previously used as a baseline for comparing the accuracy of algorithms~\cite{Leimkuhler2013Robust}.
The GJF algorithm, known for its robust performance with respect to the time step, is also included.
For the implementation of GJF method, we adopt the revised version of the algorithm proposed by Grønbech-Jensen~\cite{Grønbech-Jensen2023}, which explicitly treats the half-step velocity.
The temperature is calculated using the half-step velocity ($v_\mathrm{half}$ in the algorithm described in \ref{sec:algorithms}) for this case.

Each thermostat has specific parameters to control the relaxation toward the target temperature:
the thermostat masses $Q_1$ and $Q_2$ for the Nosé--Hoover thermostats, the relaxation time constant $\tau$ for the Bussi thermostat, and the friction coefficient $\gamma$ for the Langevin methods (see \ref{sec:algorithms} for details).
These parameter values are indicated in the corresponding figure legends and captions.
For the NHC2 thermostat we set $Q_2 = Q_1$.

We have also performed the Monte Carlo (MC) simulation as a firm standing point to sample configurations from the canonical ensemble~\cite{Frenkel_Smit}.
In the simulation, we randomly choose a particle $i$ and propose a displacement for the position of the particle $i$: $\bm{r}_i + \delta \bm{r}$.
The displacement vector $\delta \bm{r}$ is randomly drawn from the cubic box with the linear size $\delta = 0.15$~\cite{Berthier_2007}.
Then this proposal is accepted with the Metropolis condition.
For MC, we defined $N$ trials as 1 MC step, and this is defined as a time unit.

\subsection{Simulation procedure}
We first equilibrate the system at a temperature of $T = 2.0$ via an \textit{NVT} simulation using the NHC1 thermostat.
The target temperature is then changed to $T = 1.0$, and another \textit{NVT} simulation is conducted to reach equilibrium at the new temperature.
The same procedure is also carried out for $T = 0.5$.
These simulations are repeated for 320 different initial conditions to ensure good statistics.
The equilibration runs are performed with a time step of $\Delta t = 0.005$.
Using these equilibrated configurations as initial conditions, we subsequently perform production runs for each thermostat.
During the production runs, instantaneous configurations are saved at intervals of $4\tau_\alpha$.
The structural relaxation time $\tau_\alpha$ is defined via the self part of the intermediate scattering function as $F_s(k, \tau_\alpha) = e^{-1}$, where $k = 7.25$ is set to the position of the first peak of the static structure factor~\cite{Kob_Andersen_I_1995}.
For MD, $\tau_\alpha \approx 1.24$ at $T = 1.0$ and $\tau_\alpha \approx 1.17 \times 10^2$ at $T = 0.5$.
For MC, $\tau_\alpha \approx 8.46 \times 10^2$ at $T = 1.0$ and $\tau_\alpha \approx 6.87 \times 10^4$ at $T = 0.5$.
The relaxation times for MD and MC are expressed in LJ time units and MC steps, respectively; see above for definitions.
Note that $T = 1.0$ corresponds to the onset temperature of the slow relaxation, while $T = 0.5$ represents a mildly supercooled condition~\cite{Sastry_1998}.
Each production run is carried out for a total duration of $4000\tau_\alpha$.
Accordingly, each histogram in the following sections is constructed from a dataset comprising \num{320000} data points.

We use our in-house code for all simulations.
We have also performed simulations with the LAMMPS package~\cite{THOMPSON2022108171} and quantitatively checked the validity of our results.

\section{Temperature}
\label{sec:temperature}
The first observable we examine to evaluate the performance of each thermostat is the temperature $T$, which is calculated using the equipartition law
\begin{align}
\frac{\NDOF}{2} k_B T = \frac{1}{2} \sum_{i=1}^N m v_i^2,
\end{align}
where $\NDOF = dN$ denotes the total number of degrees of freedom in the system ($d = 3$ is the spatial dimension).
This relation implies that the velocity distribution of particles should follow the Maxwell--Boltzmann distribution
\begin{align}
f(v_i) = \pqty{\frac{m}{2 \pi k_B T}}^{3/2} 4\pi v_i^2 \exp\pqty{-\frac{m v_i^2}{2 k_B T}}
\end{align}
which arises from the canonical ensemble.
Accordingly, we begin by analysing whether this distribution is realised in the trajectories generated by each thermostat.

\begin{figure}
\centering
\includegraphics[width=.6\linewidth]{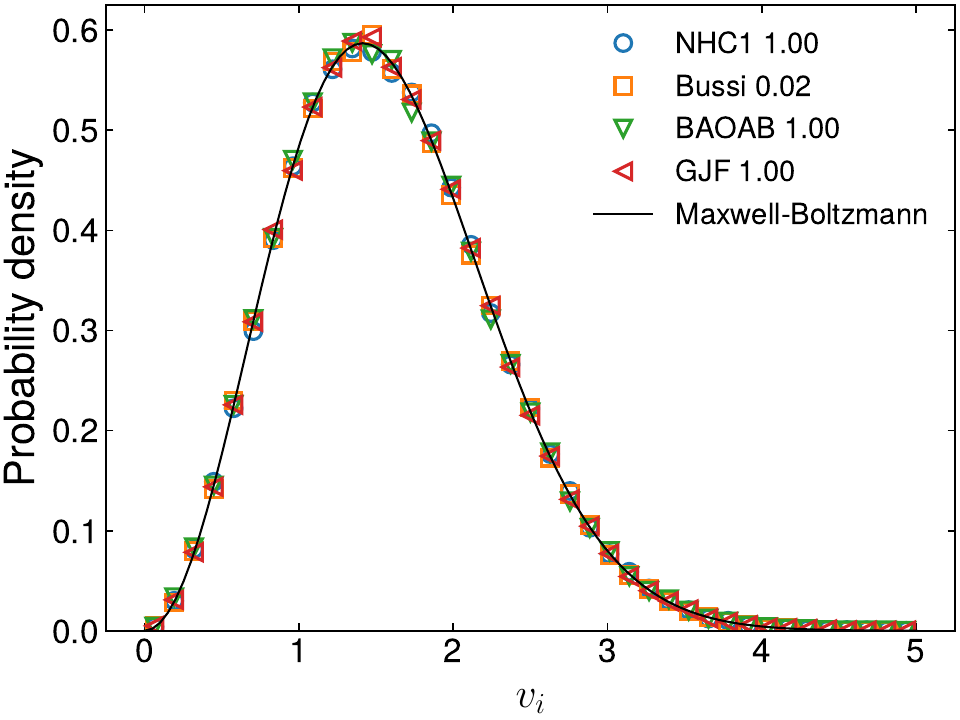}
\caption{Velocity distribution of particles in configurations at $T = 1.0$, compared with the Maxwell--Boltzmann distribution (black curve).
The parameters used for each thermostat are indicated in the legend.}
\label{fig:MB}
\end{figure}

Figure~\ref{fig:MB} shows the empirical velocity distribution of particles for $T = 1.0$, compared against the theoretical Maxwell--Boltzmann curve.
To obtain sufficient statistics, we sampled 100 snapshots from a trajectory in the production run and aggregated all particle velocities, resulting in \num{100000} data points.
Results are shown for the NHC1 and Bussi thermostats, along with two implementations of the Langevin thermostat (BAOAB and GJF algorithms), all with time step $\Delta t = 0.005$.
As shown, all the thermostats reproduce the expected distribution well.
We also confirmed this agreement for smaller time steps, such as $\Delta t = 0.001$.

\begin{figure}
\centering
\includegraphics[width=.6\linewidth]{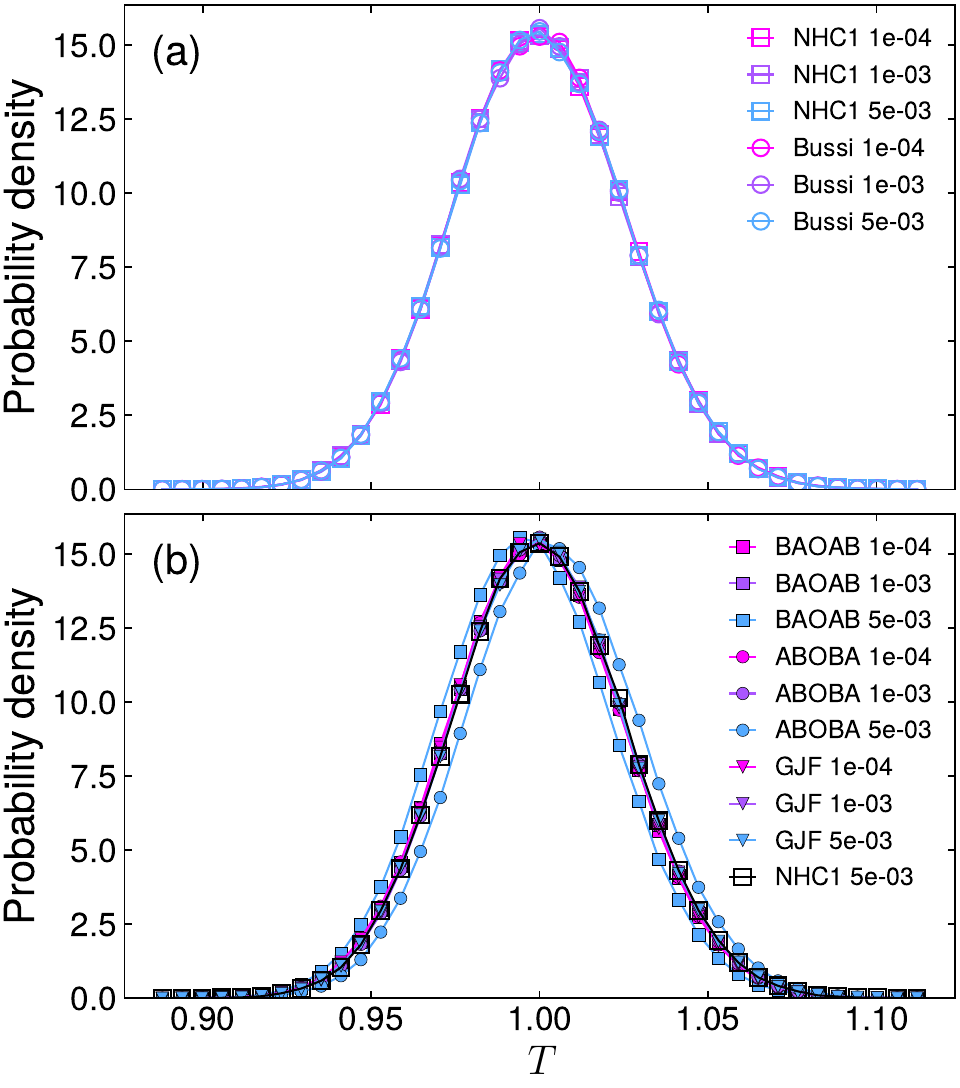}
\caption{Probability distribution of instantaneous temperature from sampled configurations at $T = 1.0$.
Panel (a) shows the results for the NHC1 and Bussi thermostats at $\Delta t = 0.005$, $0.001$, and $0.0001$, as indicated in the legend.
Panel (b) shows the results for the Langevin thermostats using the BAOAB, ABOBA, and GJF algorithms at the same time steps.
The NHC1 result at $\Delta t = 0.005$ is included as a reference.
The parameters of each thermostat are: $Q = 1.0$ for NHC1, $\tau = 0.02$ for Bussi, and $\gamma = 25.0$ for Langevin methods.}
\label{fig:histT}
\end{figure}

Next, we analyse the fluctuations of the instantaneous temperature.
The temperature is computed for each snapshot of the production trajectories, and its distribution is shown in Fig.~\ref{fig:histT}.
Panel (a) presents results for the NHC1 and Bussi thermostats, while panel (b) shows those for the Langevin thermostat, all at $\Delta t = 0.005$, 0.001, and 0.0001.
Both the NHC1 and Bussi thermostats yield robust and consistent temperature distributions across all time steps (Fig.~\ref{fig:histT} (a)).
In contrast, the Langevin thermostat exhibits a pronounced dependence on the time step (Fig.~\ref{fig:histT} (b)), as illustrated by three integration schemes: BAOAB, ABOBA, and GJF.
For reference, the distribution obtained with NHC1 at $\Delta t = 0.005$ is also included in panel (b).

We observe that BAOAB and ABOBA exhibit significant deviations at larger time steps.
At $\Delta t = 0.005$, the peaks of their distributions are noticeably shifted relative to that of the NHC1 thermostat, although the direction of deviation differs between the two.
As the time step is reduced, both distributions converge towards the NHC1 result, indicating that the latter provides a reliable reference.

In contrast, the GJF algorithm yields a robust temperature distribution that remains consistent even at the largest time step considered.
This observation is consistent with theoretical results by Grønbech-Jensen and coworker~\cite{2GJ2019,GJ2020}, who demonstrated that the use of half-step velocities enables accurate sampling of the canonical distribution, even in simple yet realistic systems such as Lennard-Jones liquids.

\begin{figure}
\centering
\includegraphics[width=.6\linewidth]{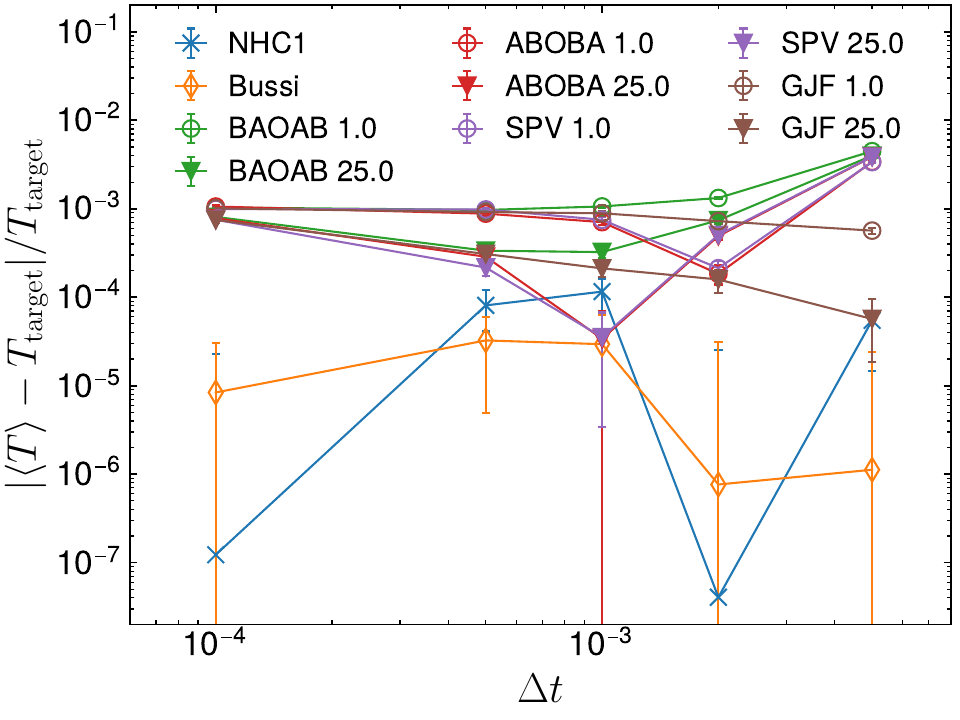}
\caption{Relative error of the ensemble-averaged temperature with respect to the target value $T_\mathrm{target} = 1.0$, shown as a function of the time step $\Delta t$.
The error bars are evaluated by the bootstrap method~\cite{Newman} with the number of repetitions of 100.
In the legend, we present the friction $\gamma$ for the Langevin methods.
The mass of the NHC1 thermostat is set to $Q = 1.0$, and the time constant of the Bussi thermostat is set to $\tau = 0.02$.}
\label{fig:errorT}
\end{figure}

To quantitatively assess the deviation of the temperature from the target value, we calculate the relative error with respect to the target temperature $T_\mathrm{target}$, defined as $|\ev{T} - T_\mathrm{target}| / T_\mathrm{target}.$
As shown in Fig.~\ref{fig:errorT}, the NHC1 and Bussi thermostats exhibit remarkably low relative errors, consistently outperforming the Langevin-based algorithms.
In particular, their errors can reach magnitudes as low as \num{e-7} at certain time step values.
Presumably due to this extremely small amplitude, no clear systematic dependence on $\Delta t$ is observed, which likely reflects the high numerical precision of these thermostats.

In contrast, the Langevin thermostats generally yield larger relative errors, typically one to two orders of magnitude higher.
Nevertheless, important differences emerge among the Langevin algorithms.
Specifically, while the BAOAB, ABOBA, and SPV schemes all exhibit a noticeable increase in error at the larger time step $\Delta t = 0.005$, the half-step velocity in the GJF integrator maintains a comparably small error across all tested time steps.
Notably, its error at $\Delta t = 0.005$ remains close to that at $\Delta t = 0.0001$, indicating superior robustness.
This trend is consistent across different values of the friction coefficient $\gamma$.

\section{Potential energy}
\label{sec:potential}
\begin{figure}
\centering
\includegraphics[width=.6\linewidth]{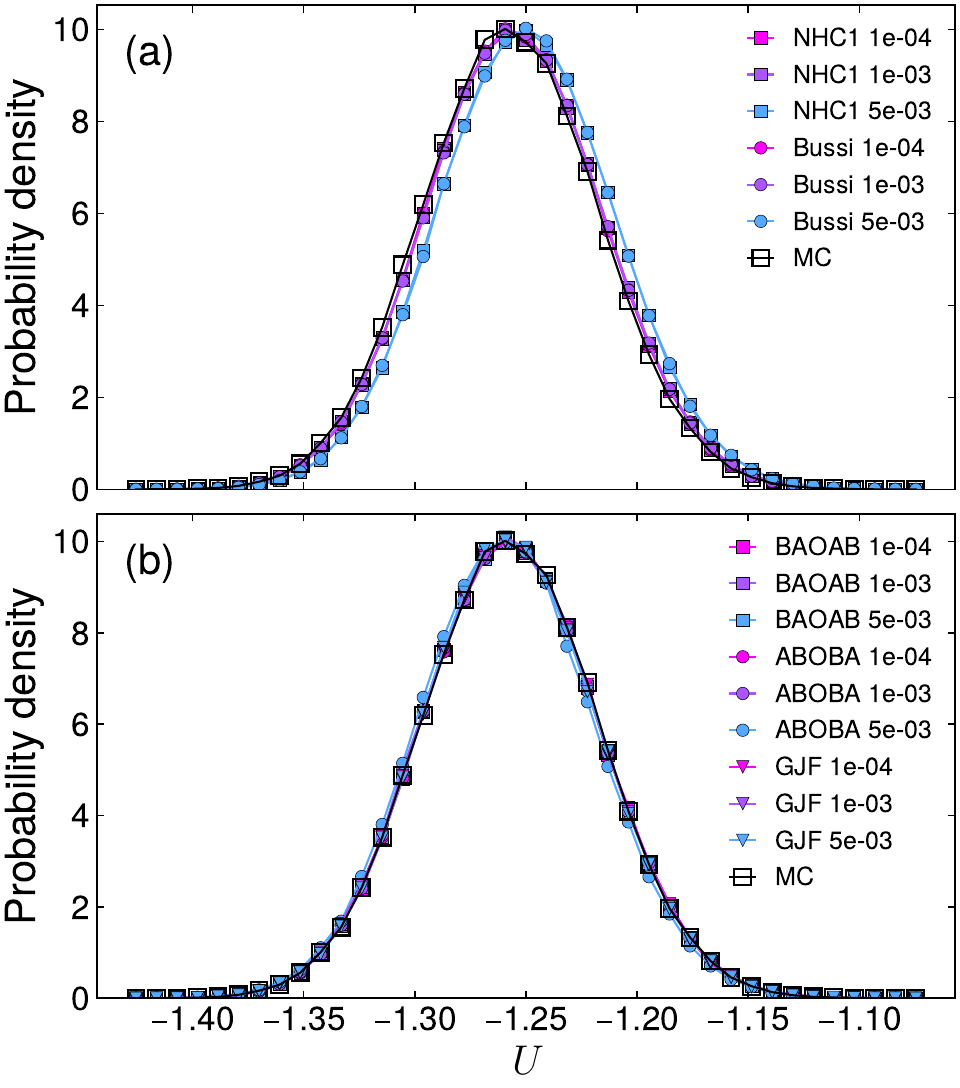}
\caption{Probability distribution of potential energy at $T = 1.0$ for (a) NHC1 and Bussi thermostats and (b) Langevin methods.
Each method presents data with $\Delta t = 0.005$, $0.001$, and $0.0001$, as indicated in the legend.
The parameters of each thermostat are: $Q = 1.0$ for NHC1, $\tau = 0.02$ for Bussi, and $\gamma = 25.0$ for Langevin methods.
The MC distribution, representing the canonical ensemble, is shown for comparison.}
\label{fig:histU}
\end{figure}

Next, we examine the distribution of potential energy.
As for the temperature case, we present in Fig.~\ref{fig:histU} (a) the distributions obtained with the NHC1 and Bussi thermostats, and in Fig.~\ref{fig:histU} (b) that obtained with the Langevin thermostat.
For reference, the potential energy distribution from MC configurations is also included, representing the canonical ensemble at the target temperature $T = 1.0$.
With sufficient statistics, the MC sampling generates configurations exactly from the canonical ensemble without integration errors; therefore, we employ its distribution as a reference.
Compared with the results for temperature, the behaviour here is strikingly different.
In this case, the Langevin algorithms yield distributions that remain close to the MC reference for all investigated time steps $\Delta t$.
By contrast, the NHC1 and Bussi thermostats display a clear deviation from the MC distribution at the largest time step, $\Delta t = 0.005$.
As $\Delta t$ decreases, their distributions converge towards the MC reference.

\begin{figure}
\centering
\includegraphics[width=.6\linewidth]{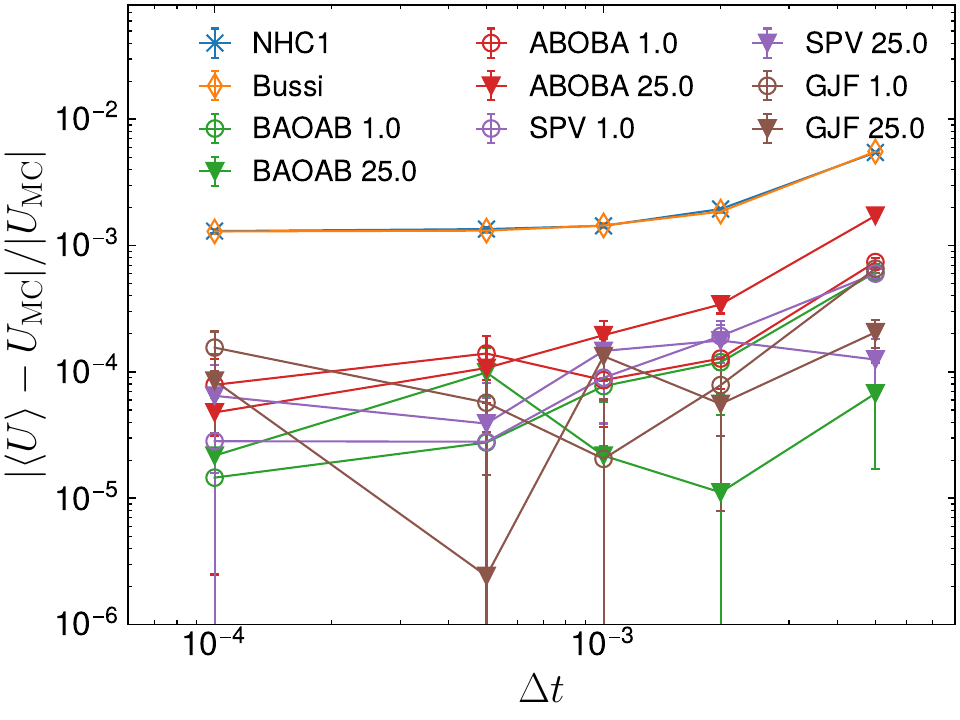}
\caption{Relative error of the ensemble-averaged potential energy with respect to the MC value $U_\mathrm{MC}$, shown as a function of the time step $\Delta t$.
The error bars are evaluated by the bootstrap method~\cite{Newman} with the number of repetitions of 100.
In the legend, we present the friction $\gamma$ for the Langevin methods.
The mass of the NHC1 thermostat is set to $Q = 1.0$, and the time constant of the Bussi thermostat is set to $\tau = 0.02$.}
\label{fig:errorU}
\end{figure}

Following the approach used for the temperature analysis, we quantify the relative error $|\ev{U} - U_\mathrm{MC}| / |U_\mathrm{MC}|$ for each thermostat (Fig.~\ref{fig:errorU}).
The NHC1 and Bussi thermostats exhibit larger errors than the Langevin algorithms for all $\Delta t$ considered.
Their errors decrease substantially when $\Delta t$ is reduced from $0.005$ to $0.001$, but then level off and do not decrease further within our range of investigation ($\Delta t \geq 0.0001$), though the converged error is approximately \SI{0.1}{\percent}, which is already small.

In contrast, the Langevin schemes show consistently smaller errors than the NHC1 and Bussi counterparts.
Although the error exhibits an increasing trend with time step, the values remain at around \SI{0.01}{\percent}.
The data are noisy, presumably due to the very small magnitude of the error.
In our case, the relative errors are on the order of \num{e-4} to \num{e-5}, which is substantially smaller than the \num{e-3} level of the alanine dipeptide protein reported by Leimkuhler and Matthews~\cite{Leimkuhler2012Rational,Leimkuhler2013Robust}.
As a consequence, we cannot infer a clear order of accuracy from the present results.

\begin{figure}
\centering
\includegraphics[width=.6\linewidth]{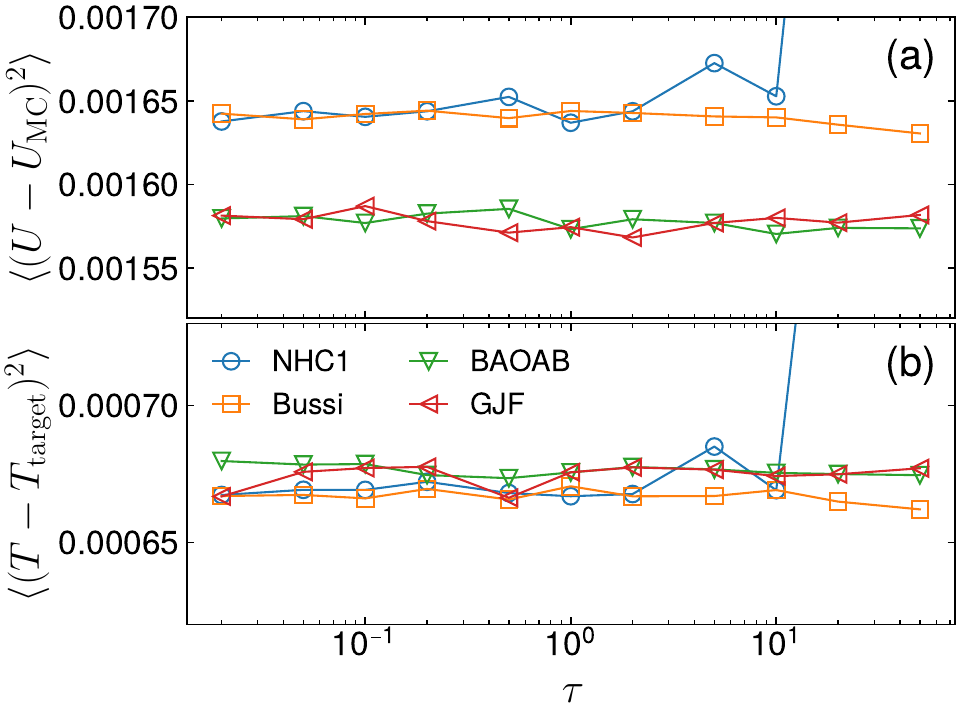}
\caption{The variance of (a) potential energy and (b) temperature of each thermostat, shown as a function of time constant $\tau$ of each thermostat.
The target temperature is $T_\mathrm{target} = 1.0$ and the integration time step is $\Delta t = 0.005$ for all methods.}
\label{fig:tau_thermo}
\end{figure}

Finally, we investigate how the temperature and potential energy distributions discussed above depend on the time constant $\tau$ of the thermostat.
Figure~\ref{fig:tau_thermo} shows the variance of the sampled distributions as a function of time constants $\tau$ spanning more than three orders of magnitude.
For the comparison between different schemes, we performed a simulation starting from a configuration at $T = 2.0$ targeted at $T = 1.0$.
We monitored the temperature during the relaxation process and confirmed that $\tau = 0.02$ for the Bussi thermostat yields a convergence time comparable to that obtained with $Q = 1.0$ for the NHC1 thermostat and $\gamma = 1.0$ for the BAOAB and GJF schemes.
We use these parameter values as reference points for comparison.
While the Nosé--Hoover thermostat exhibits a pronounced increase in variance at larger time constants, the Bussi thermostat shows a variance that is essentially independent of $\tau$, consistent with the original report~\cite{Bussi_2007}.
This robustness with respect to the time constant is directly related to efficient sampling performance.
The Langevin schemes (BAOAB and GJF) also display stable variances across the range of $\tau$ examined.
A closer inspection suggests that the Bussi thermostat yields an even more uniform variance than the Langevin schemes, although a more systematic analysis would be required to confirm this trend.

\section{Structural quantity}
\label{sec:rdf}
\begin{figure}
\centering
\includegraphics[width=.6\linewidth]{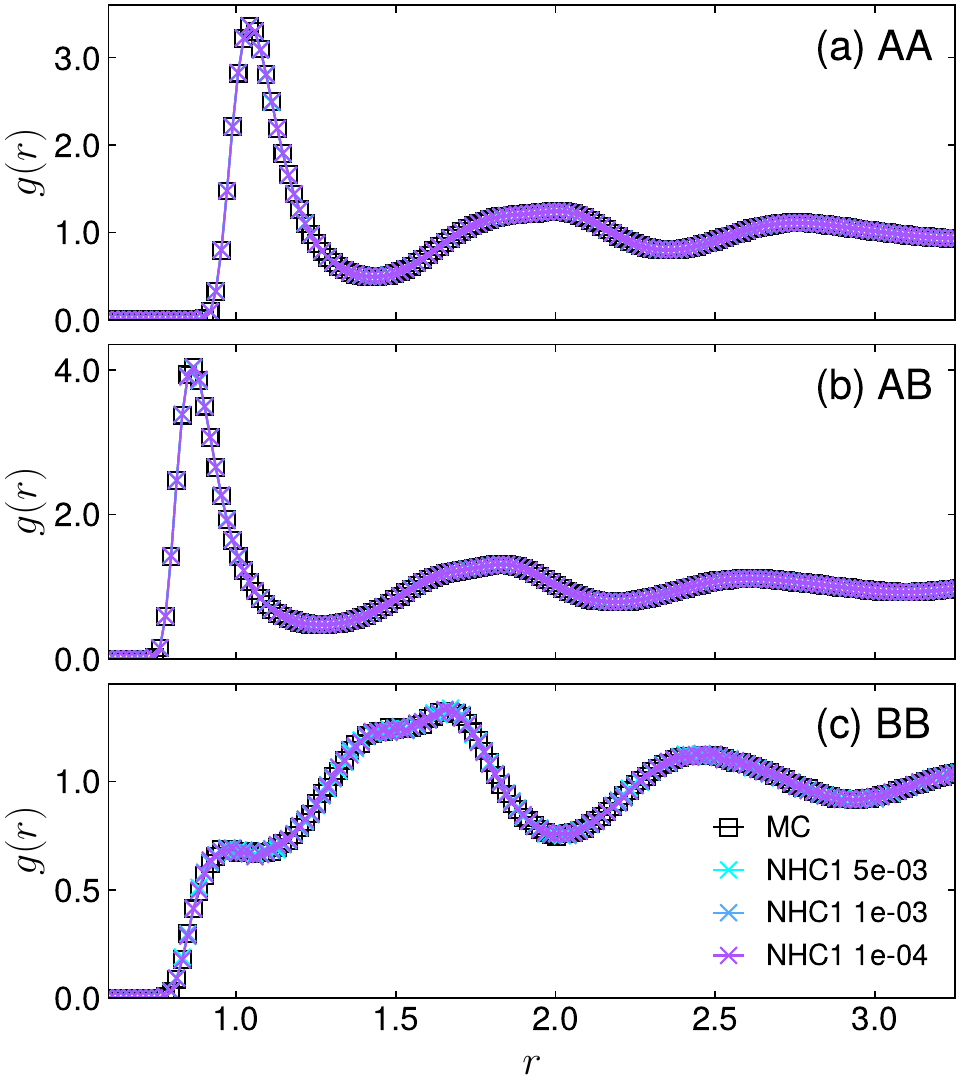}
\caption{The radial distribution function $g(r)$ calculated from configurations at temperatures $T = 1.0$.
Panels (a), (b), and (c) show AA, AB, and BB correlations, respectively.
The mass of NHC1 thermostat is set to $Q = 1.0$ and the time step $\Delta t$ of each data is indicated in the legend.}
\label{fig:RDF}
\end{figure}

The results presented in the previous section indicate that the NHC1 and Bussi thermostats exhibit a stronger $\Delta t$ dependence in the potential energy compared to the Langevin thermostats.
Since the potential energy reflects the static structure of the system, we next examine the radial distribution function (RDF) $g(r)$ to assess whether the structure is likewise affected by the choice of time step.
We compute $g(r)$ using the \texttt{freud} package~\cite{freud2020}.
Figure~\ref{fig:RDF} shows RDFs for configurations obtained with the NHC1 thermostat at various time steps: $\Delta t = 0.005$, 0.001, and 0.0001.
As a reference, we also include RDFs computed from MC configurations.
Across all particle correlations (AA, AB, and BB), the RDFs from the NHC1 configurations closely align with those from the MC reference, with no appreciable deviations observed.
This finding, taken together with the potential energy trends discussed earlier, suggests that although increasing $\Delta t$ induces a systematic shift in the potential energy, it has a negligible effect on two-point structural correlations as characterised by $g(r)$.
The consistency of $g(r)$ across time steps is reassuring; while the average potential energy drifts with $\Delta t$, the underlying structure remains essentially unchanged.
This indicates that the NHC1 thermostat functions properly as a sampling algorithm for configurations at the target temperature.

\section{Dynamics}
\label{sec:dynamics}
\begin{figure}
\centering
\includegraphics[width=.6\linewidth]{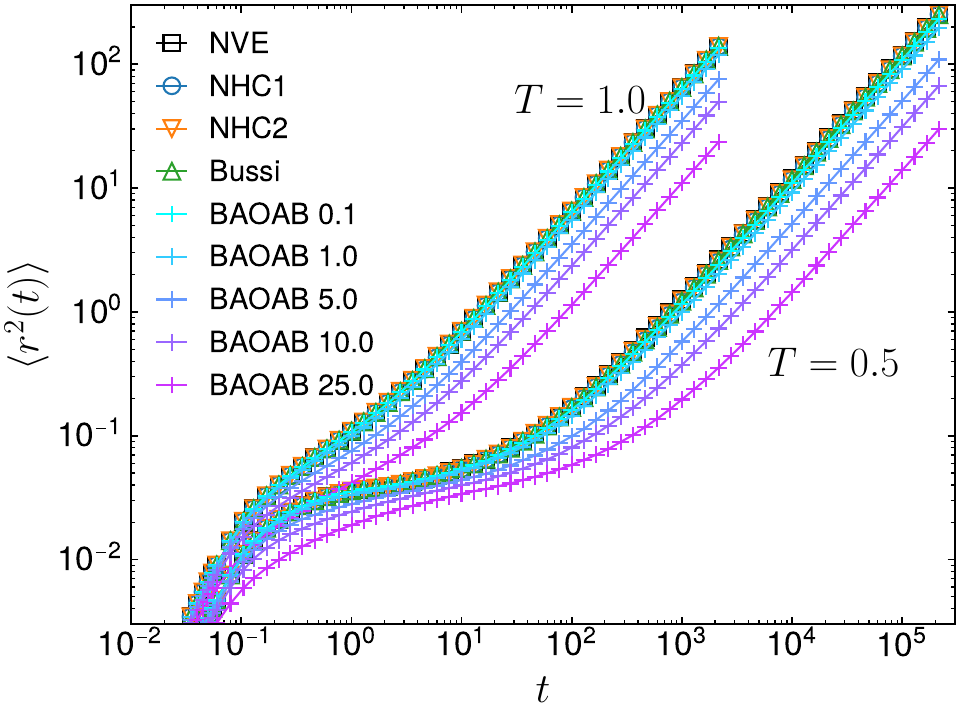}
\caption{Mean squared displacements $\ev{r^2(t)}$ of A particles at temperatures $T=1.0, 0.5$ for various integration schemes.
The thermostat parameters are: $Q_1 = Q_2 = 1.0$ for the Nosé--Hoover thermostats and $\tau = 0.02$ for the Bussi thermostat.
The friction coefficient $\gamma$ is varied from $\gamma = 0.1$ to $\gamma = 25.0$, as indicated in the legend.
The time step is set to $\Delta t = 0.005$ for all integration schemes.}
\label{fig:MSD}
\end{figure}

Finally, we investigate the relaxation dynamics under each thermostat by computing the mean squared displacement (MSD), defined as
\begin{align}
\ev{r^2(t)} = \frac{1}{N_\mathrm{A}} \sum_{i=1}^{N_\mathrm{A}} \ev{\abs{\bm{r}_i(t) - \bm{r}_i(0)}^2}.
\end{align}
Figure~\ref{fig:MSD} presents the MSD at two temperatures, $T = 1.0$ and $T = 0.5$, for each thermostat, alongside results obtained using the velocity Verlet integrator~\cite{Swope1982} in its standard `kick-drift-kick' form (labelled as NVE).
At both temperatures, the Nosé--Hoover thermostats (NHC1 and NHC2) and the Bussi thermostat produce MSDs that are in excellent agreement with those obtained from velocity Verlet integration, indicating that these thermostats are in close agreement with the Newtonian dynamics.

In contrast, the Langevin thermostat yields markedly different relaxation behaviour.
Here, we focus on the BAOAB splitting scheme for illustrative purposes, although we have verified that other integrator variants (ABOBA, SPV, and GJF) produce consistent results, as presented in \ref{sec:MSD_Langevin}.
The most notable difference is that the MSD deviates systematically from that of Newtonian dynamics.
As the friction coefficient $\gamma$ increases, the long-time MSD decreases, implying a suppression of the diffusion coefficient.
Furthermore, the Langevin dynamics also affects the shape of the MSD near the plateau, namely, the $\beta$-relaxation regime in the language of glass transition physics.
While the Newtonian dynamics and specific thermostats exhibits sharp change between ballistic and plateau regimes and between plateau and diffusive regimes, the Langevin dynamics exhibits a more gradual transition into the plateau~\cite{Gleim_1998}, particularly at higher $\gamma$.
The plateau is no longer flat in the Langevin dynamics, especially at higher $\gamma$, at least in our temperature range.
Notably, the short-time relaxation behaviour under the Langevin thermostat with stronger frictions resembles that observed in MC dynamics~\cite{Berthier_2007}.

\begin{figure}
\centering
\includegraphics[width=.6\linewidth]{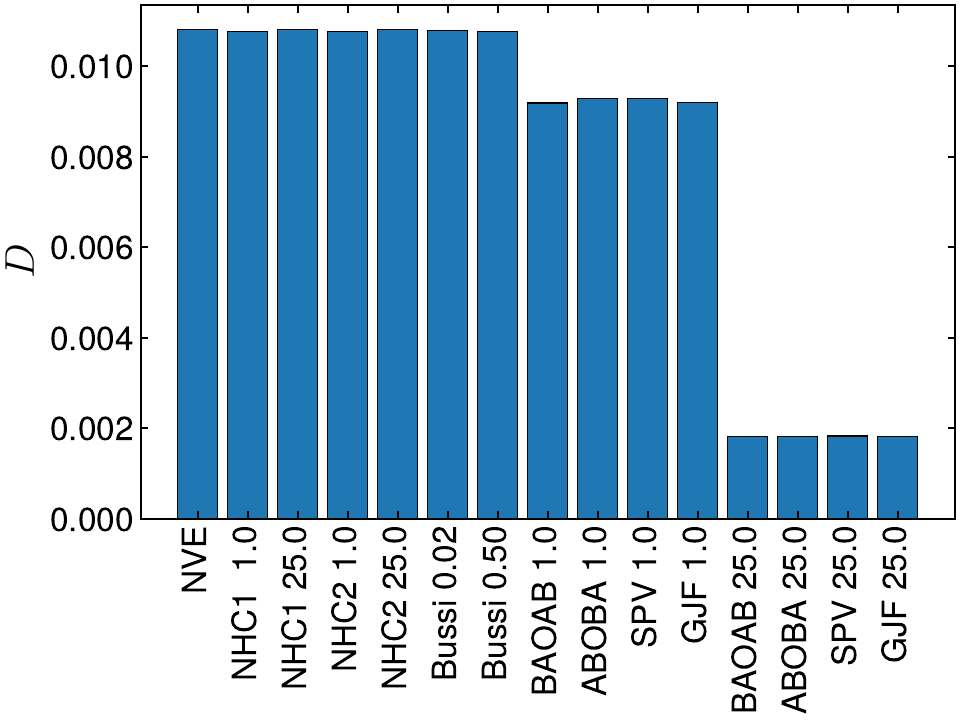}
\caption{Diffusion coefficient of different thermostats at $T = 1.0$.
Thermostat parameters are given in the label with each thermostat name.}
\label{fig:diffusion}
\end{figure}

\begin{figure}
\centering
\includegraphics[width=.6\linewidth]{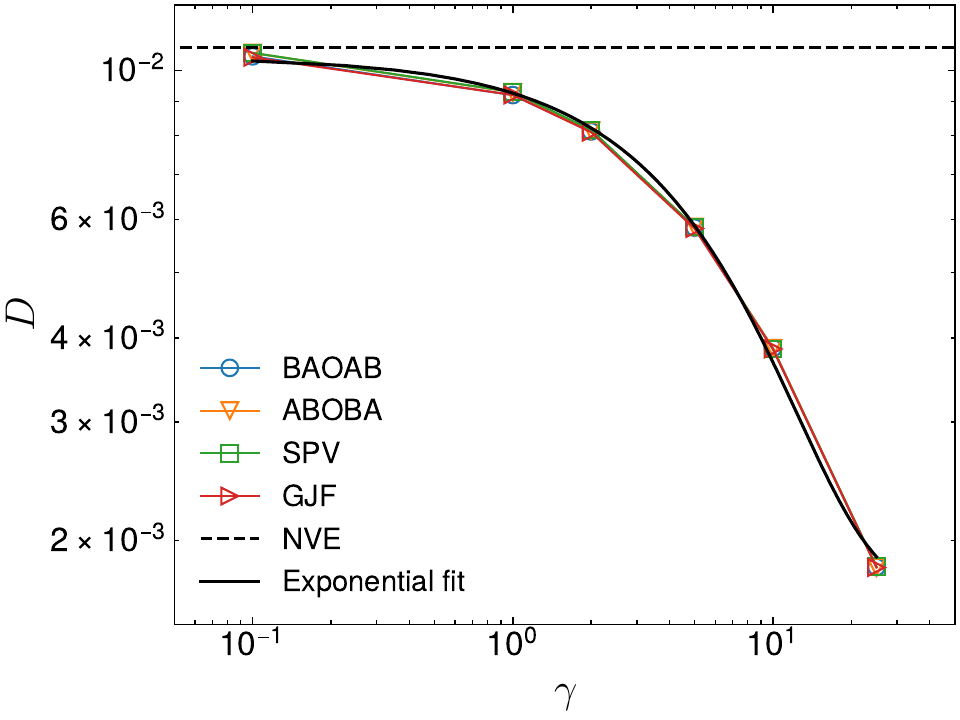}
\caption{Diffusion coefficient of the Langevin thermostats at $T = 1.0$.
The dashed horizontal line represents the value of the NVE case.
The fitted curve $A\exp\pqty{-B\gamma} + C$, where $A = 0.00878401$, $B = 0.14747297$, and $C = 0.00167097$, is also included.}
\label{fig:diffusion_Langevin}
\end{figure}

We next calculate the diffusion coefficient $D = \lim_{t \to \infty} \ev{r^2(t)}/6t$ to quantify the extent to which the Langevin thermostat suppresses diffusion compared to the NVE case.
Figure~\ref{fig:diffusion} presents the diffusion coefficient $D$ obtained using each thermostat.
As readily observed in the MSD, the Nosé--Hoover and Bussi thermostats yield diffusion coefficients similar to that of the NVE case.
On the other hand, the Langevin thermostat leads to a marked reduction in $D$ as $\gamma$ increases.
The deterioration of $D$ is fitted with an exponential curve $A\exp\pqty{-B\gamma} + C$ as illustrated in Fig.~\ref{fig:diffusion_Langevin}, which shows the best-fit exponential curve.

\section{Low-temperature results}
\label{sec:lowT}
Before we wrap up this paper, we present the results of the low-temperature simulations at $T = 0.5$.
For the binary Lennard-Jones model under the current investigation, this temperature is located in the supercooled regime, where the time correlation function exhibits a plateau and the famous two-step relaxation emerges~\cite{Kob_Andersen_I_1995}.
In addition to the dynamical behaviour, the physics behind the glass transition phenomenology is more affected by the topographic nature of the complex potential energy landscape~\cite{Sastry_1998,Debenedetti_2001}.

\begin{figure}
\centering
\includegraphics[width=.6\linewidth]{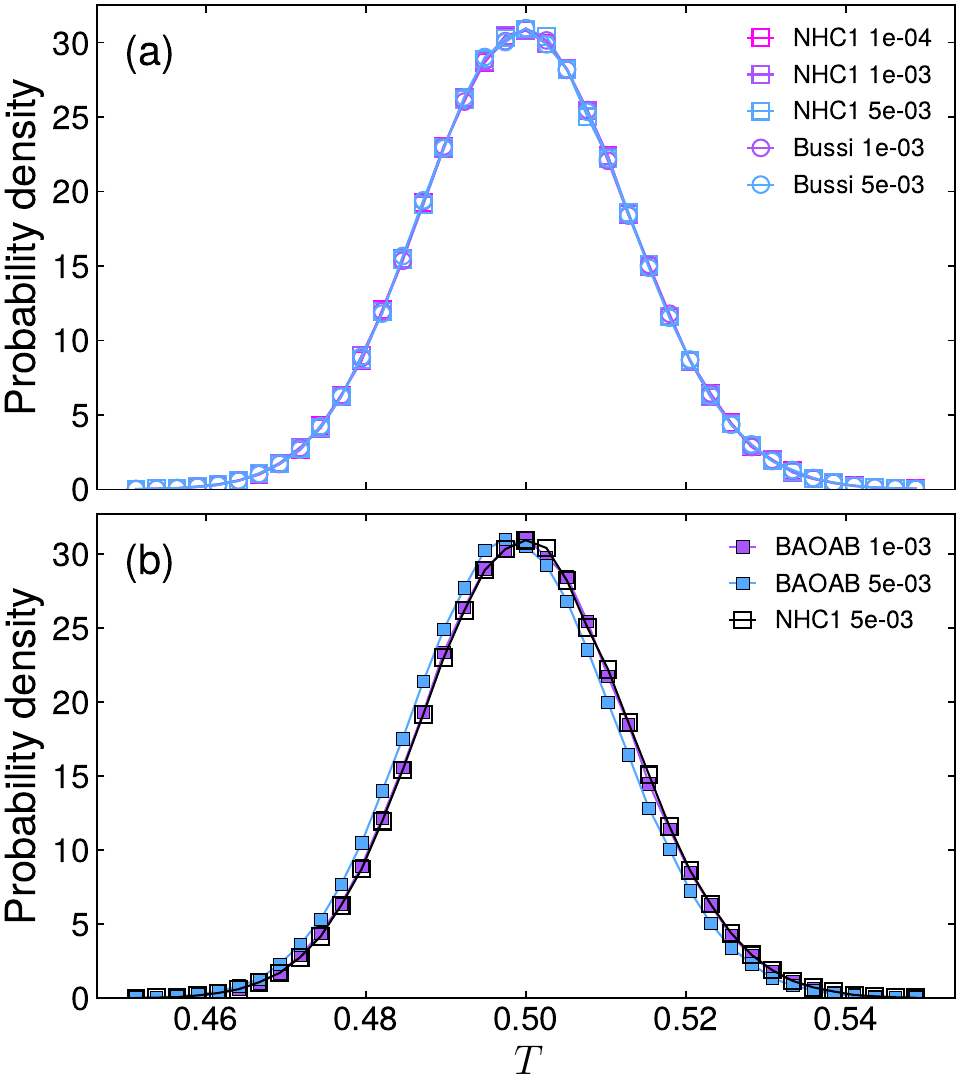}
\caption{Probability distribution of instantaneous temperature from sampled configurations at $T = 0.5$.
Panel (a) shows the results for the NHC1 thermostat at $\Delta t = 0.005$, $0.001$, and $0.0001$, and for the Bussi thermostat at $\Delta t = 0.005, 0.001$, as indicated in the legend.
Panel (b) shows the results for the BAOAB scheme at $\Delta t = 0.005, 0.001$.
The NHC1 result at $\Delta t = 0.005$ is included as a reference.
The parameters of each thermostat are: $Q = 1.0$ for NHC1, $\tau = 0.02$ for Bussi, and $\gamma = 25.0$ for Langevin method.}
\label{fig:histT_low}
\end{figure}

In Fig.~\ref{fig:histT_low}, the probability distribution for $T = 0.5$ is shown with the NHC1 thermostat in the panel (a) and the Langevin thermostat with the BAOAB splitting in the panel (b).
Consistent with the results with $T = 1.0$ (Fig.~\ref{fig:histT}), the NHC1 thermostat yields a very precise distribution with varying time step $\Delta t$, and the Langevin thermostat suffers from the visible error at the large $\Delta t$.

\begin{figure}
\centering
\includegraphics[width=.6\linewidth]{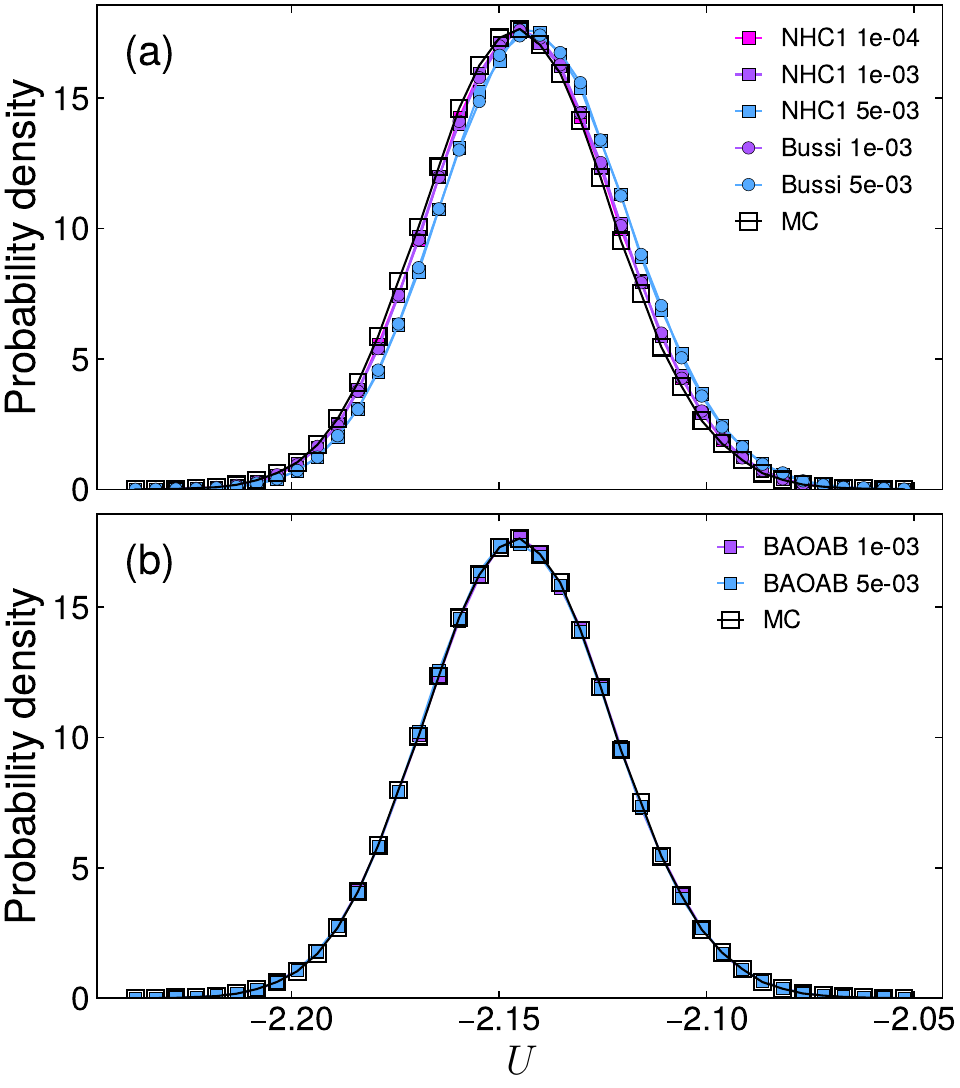}
\caption{Probability distribution of potential energy at $T = 0.5$ for (a) NHC1 and Bussi thermostats and (b) BAOAB scheme.
Each method presents data with different $\Delta t$, as indicated in the legend.
The parameters of each thermostat are: $Q = 1.0$ for NHC1, $\tau = 0.02$ for Bussi, and $\gamma = 25.0$ for Langevin methods.
The MC distribution, representing the canonical ensemble, is shown for comparison.}
\label{fig:histU_low}
\end{figure}

On the other hand, in the probability distribution of potential energy (Fig.~\ref{fig:histU_low}), the NHC1 thermostat exhibits a deviation from the MC distribution.
The Langevin thermostat again shows the robust distribution.
The results in this section prove the consistency of our results obtained at $T = 1.0$ for lower temperatures in the supercooled regime.

\section{Conclusions and final remarks}
\label{sec:conclusion}
We have characterised the sampling behaviour of seven thermostat algorithms in the MD simulation of binary Lennard-Jones particles.
A key finding from our study is that these methods can be broadly divided into two categories regarding how the error scales with the time step $\Delta t$.

The first category comprises the Nosé--Hoover chain and Bussi thermostats.
These thermostats yield highly accurate temperature distributions but show a small yet noticeable deviation in the potential energy distribution.
Although the RDF is indistinguishable from that obtained from MC configurations, this deviation could introduce artefacts in properties requiring precise potential energy fluctuations, such as the specific heat.

This behaviour can be understood by considering the discretisation error and the coupling introduced by the thermostat.
In symplectic integration, the conserved quantity during the time evolution is not the true Hamiltonian of the system, but a perturbed quantity usually referred to as the shadow Hamiltonian~\cite{Yoshida1990}.
Although the exact form of the shadow Hamiltonian cannot be obtained for general systems, its deviation from the true Hamiltonian has the same order in $\Delta t$ as the integration scheme.
For the Nosé--Hoover chain thermostat, the shadow Hamiltonian of the extended system, including both the target system and the thermostat variables, is conserved by the symplectic integration.
Because this method rescales particle velocities via the thermostat variables, the system's temperature of the target system tends to approach the true canonical value.
Consequently, the finite discretisation error is more likely to manifest in the potential energy, leading to deviations at larger $\Delta t$.
In the Bussi thermostat, particle velocities are rescaled simultaneously with a stochastic factor, so a strict shadow Hamiltonian does not exist.
However, its integration scheme is still largely equivalent to the velocity Verlet algorithm, which is a second-order symplectic integrator.
As with the Nosé--Hoover chain, the thermostat directly controls the velocities, maintaining the temperature near its canonical value, while residual discretisation errors predominantly manifest in the potential energy.

In contrast, Langevin thermostats exhibit the opposite tendency:
their potential energy distributions closely match those from MC even at larger $\Delta t$, but temperature sampling shows a stronger $\Delta t$-dependence.
This behaviour reflects the fact that the stochastic noise only couples to the velocities.
In contrast, the half-step velocity in the GJF method maintains a robust temperature distribution across $\Delta t$ values.
For example, in simulations targeting $T_\mathrm{target} = 1.0$ with $\Delta t = 0.005$, the NHC1 and Bussi thermostats give $\ev{T} = 1.0000547$ and $\ev{T} = 1.0000011$, respectively, whereas the Langevin thermostat with the BAOAB splitting yields $\ev{T} = 0.9960139$.
Reducing $\Delta t$ alleviates this discrepancy;
at $\Delta t = 0.0001$, BAOAB recovers $\ev{T} = 0.9991907$.
Amongst the Langevin schemes, the GJF method performs particularly well, achieving $\ev{T} = 0.9999432$ even at $\Delta t = 0.005$, thus providing a favourable balance between potential and temperature sampling accuracy.

From a practical standpoint, the choice of thermostat may therefore depend on the property of interest:
Nosé--Hoover or Bussi thermostats are preferable when accurate temperature control is paramount, while Langevin schemes, especially GJF, are advantageous for accurately sampling potential energy distributions at moderate $\Delta t$.

\begin{figure}
\centering
\includegraphics[width=.6\linewidth]{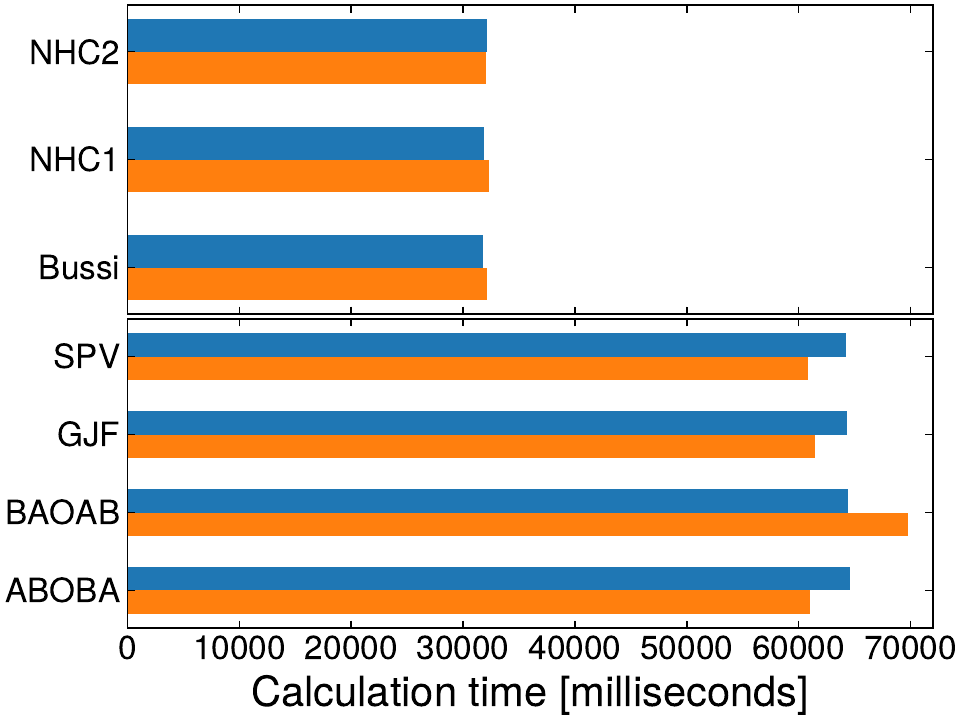}
\caption{CPU time required to perform MD simulations for $t = 500.0$ using different thermostat schemes with two different parameters.
The upper blue data in each entry corresponds to $\tau = 0.02$ of Bussi thermostat ($Q = 1.0$, $\gamma = 1.0$), while the lower orange data corresponds to $\tau = 0.50$ of Bussi thermostat ($Q = 25.0$, $\gamma = 25.0$).
The CPU times are measured for ten repetitions following a test run.}
\label{fig:cputime}
\end{figure}

We compare the computational cost of each thermostat for a fixed physical simulation time (Fig.~\ref{fig:cputime}).
All runs were performed on a laboratory-level computing node equipped with an Intel Xeon E5-2683 v4 CPU, compiled with the Intel oneAPI C++ Compiler 2021.4.0 using \texttt{-O3 -xCORE\_AVX2} optimisation flags.
The Langevin simulations are approximately twice as slow as Nosé--Hoover or Bussi runs.
This is primarily because the Langevin thermostat requires $\NDOF$ independent normal random numbers at every MD increment, whereas the deterministic Nosé--Hoover method does not.
The Bussi thermostat also uses normal random numbers;
however, due to its global thermostat nature, these numbers appear as a squared sum in the algorithm, allowing them to be reduced to a single Gamma random number~\cite{Bussi_2007}.
This reduces its computational cost compared with the Langevin schemes, which act as local thermostats, as noted in the Introduction.

Finally, we note that these conclusions are drawn from a binary Lennard-Jones system in the \textit{NVT} ensemble.
While the trends are expected to be general, further testing on larger or more complex molecular systems would be a natural extension.
Nonetheless, the present results provide a straightforward reference for practitioners to select thermostat schemes balancing accuracy in target observables with computational efficiency.

\section*{CRediT authorship contribution statement}
\textbf{K.~Shiraishi:} Conceptualization, Data curation, Formal analysis, Investigation, Methodology, Project administration, Software, Validation, Visualization, Writing - original draft, Writing - review \& editing;
\textbf{E.~Minamitani:} Conceptualization, Funding acquisition, Methodology, Project administration, Resources, Supervision, Validation, Writing - review \& editing;
\textbf{K.~Kim:} Conceptualization, Formal analysis, Funding acquisition, Methodology, Project administration, Supervision, Validation, Writing - review \& editing.

\section*{Data availability}
Data relevant to this work can be accessed at the Zenodo repository~\cite{Zenodo}.

\section*{Declaration of competing interest}
The authors declare that they have no known competing financial interests or personal relationships that could have appeared to influence the work reported in this paper.

\section*{Acknowledgements}
We thank Kento Kasahara for helpful discussions.
This work is supported by JSPS KAKENHI Grant Numbers JP22K03550, JP24H01719, JP25K00968, and JP25H01478.

\appendix
\section{Mean squared displacements of Langevin thermostats}
\label{sec:MSD_Langevin}

\begin{figure}
\centering
\includegraphics[width=\linewidth]{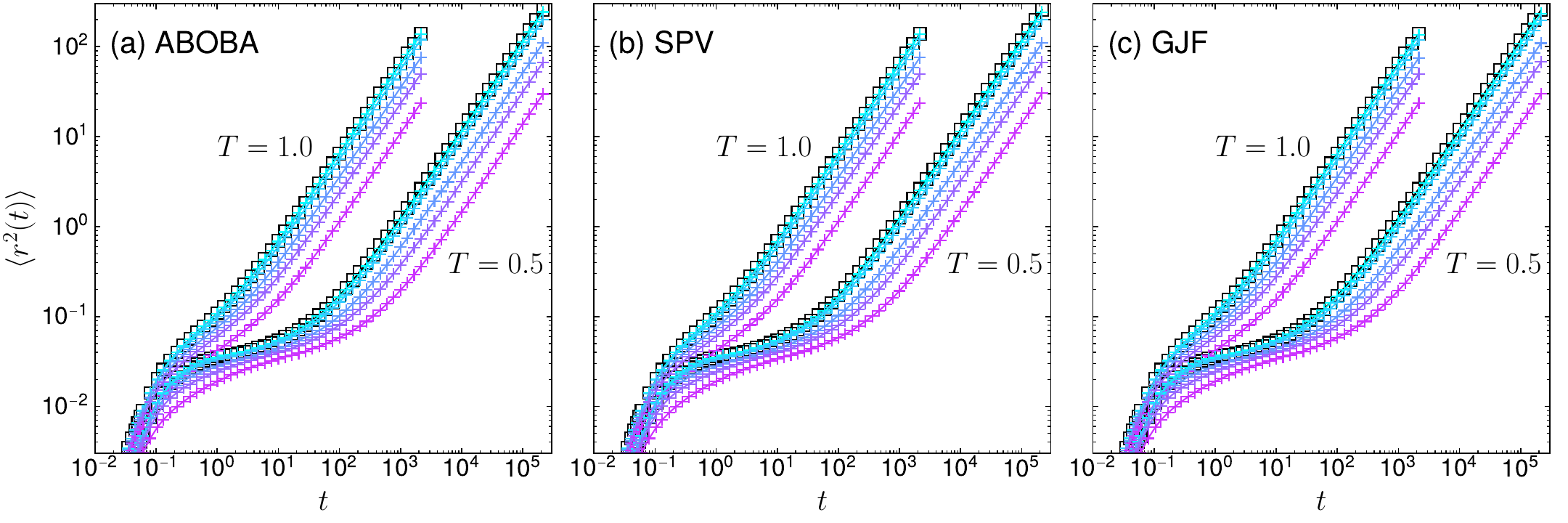}
\caption{Mean squared displacements $\ev{r^2(t)}$ of A particles at temperatures $T=1.0$ and $0.5$, obtained with three Langevin thermostats.
Panel (a), (b), and (c) correspond to the ABOBA, SPV, and GJF methods, respectively.
The legend is identical to that of Fig.~\ref{fig:MSD}.}
\label{fig:msd_langevin}
\end{figure}

Figure~\ref{fig:msd_langevin} shows the MSDs obtained with three Langevin thermostats (ABOBA, SPV, and GJF).
All these methods yield results essentially indistinguishable from those of the BAOAB method presented in Fig.~\ref{fig:MSD}.

\section{Algorithms}
\label{sec:algorithms}

\subsection{Nosé--Hoover}
\begin{enumerate}
 \item NHC1
       \begin{enumerate}
        \item $v_\xi \leftarrow v_\xi + \frac{\Delta t}{4} (mv^2 - d N T_\mathrm{target}) / Q$
        \item $v \leftarrow v \exp \pqty{-v_\xi \frac{\Delta t}{2}}$
        \item $v_\xi \leftarrow v_\xi + \frac{\Delta t}{4} (mv^2 - d N T_\mathrm{target}) / Q$
       \end{enumerate}
 \item $v \leftarrow v + \frac{\Delta t}{2m}F$
 \item $x \leftarrow x + \Delta t v$
 \item $v \leftarrow v + \frac{\Delta t}{2m}F$
 \item Repeat step \#1
\end{enumerate}

\subsection{Nosé--Hoover chain $M=2$}
\begin{enumerate}
 \item NHC2
       \begin{enumerate}
        \item $v_{\xi,2} \leftarrow v_{\xi,2} + \frac{\Delta t}{4} (Q_1 v_{\xi,1}^2 - T_\mathrm{target}) / Q_2$
        \item $v_{\xi,1} \leftarrow v_{\xi,1} \exp \pqty{-v_{\xi,2} \frac{\Delta t}{8}}$
        \item $v_{\xi,1} \leftarrow v_{\xi,1} + \frac{\Delta t}{4} (mv^2 - d N T_\mathrm{target}) / Q_1$
        \item $v_{\xi,1} \leftarrow v_{\xi,1} \exp \pqty{-v_{\xi,2} \frac{\Delta t}{8}}$
        \item $v \leftarrow v \exp \pqty{-v_{\xi,1} \frac{\Delta t}{2}}$
        \item $v_{\xi,1} \leftarrow v_{\xi,1} \exp \pqty{-v_{\xi,2} \frac{\Delta t}{8}}$
        \item $v_{\xi,1} \leftarrow v_{\xi,1} + \frac{\Delta t}{4} (mv^2 - d N T_\mathrm{target}) / Q_1$
        \item $v_{\xi,1} \leftarrow v_{\xi,1} \exp \pqty{-v_{\xi,2} \frac{\Delta t}{8}}$
        \item $v_{\xi,2} \leftarrow v_{\xi,2} + \frac{\Delta t}{4} (Q_1 v_{\xi,1}^2 - T_\mathrm{target}) / Q_2$
       \end{enumerate}
 \item $v \leftarrow v + \frac{\Delta t}{2m}F$
 \item $x \leftarrow x + \Delta t v$
 \item $v \leftarrow v + \frac{\Delta t}{2m}F$
 \item Repeat step \#1
\end{enumerate}

\subsection{Bussi thermostat}
\begin{enumerate}
 \item $v \leftarrow v + \frac{\Delta t}{2m}F$
 \item $x \leftarrow x + \Delta t v$
 \item $v \leftarrow v + \frac{\Delta t}{2m}F$
 \item $\alpha^2 = \exp \pqty{-\frac{\Delta t}{\tau}} + \bqty{1 - \exp \pqty{-\frac{\Delta t}{\tau}}} \frac{T_\mathrm{target}}{v^2} \pqty{n_1^2 + n_2^2 + g} + 2n_1\sqrt{\exp \pqty{-\frac{\Delta t}{\tau}} \bqty{1 - \exp \pqty{-\frac{\Delta t}{\tau}}} \frac{T_\mathrm{target}}{v^2}}$
 \item $v \leftarrow \alpha v$
\end{enumerate}

$n_1, n_2$ are random numbers sampled from a normal distribution:
$n_1, n_2 \sim \mathcal{N}(0,1)$.
$g$ is a random number sampled from the following Gamma distribution (when $\NDOF$ in the system is even):
$g \sim \mathrm{Gamma}(\frac{dN-2}{2}, 2)$.

\subsection{Langevin thermostat BAOAB}
\begin{enumerate}
 \item $v \leftarrow v + \frac{\Delta t}{2m}F$
 \item $x \leftarrow x + \frac{\Delta t}{2} v$
 \item $v \leftarrow \exp \pqty{-\gamma \Delta t} v + \sqrt{m} \sqrt{T_\mathrm{target} \bqty{1 - \exp \pqty{-2\gamma \Delta t}}} n$
 \item $x \leftarrow x + \frac{\Delta t}{2} v$
 \item $v \leftarrow v + \frac{\Delta t}{2m}F$
\end{enumerate}

$n$ is a $dN$-size vector whose elements are drawn from $\mathcal{N}(0,1)$.

\subsection{Langevin thermostat ABOBA}
\begin{enumerate}
 \item $x \leftarrow x + \frac{\Delta t}{2} v$
 \item $v \leftarrow v + \frac{\Delta t}{2m}F$
 \item $v \leftarrow \exp \pqty{-\gamma \Delta t} v + \sqrt{m} \sqrt{T_\mathrm{target} \bqty{1 - \exp \pqty{-2\gamma \Delta t}}} n$
 \item $v \leftarrow v + \frac{\Delta t}{2m}F$
 \item $x \leftarrow x + \frac{\Delta t}{2} v$
\end{enumerate}

\subsection{Langevin thermostat SPV}
\begin{enumerate}
 \item $x \leftarrow x + \frac{\Delta t}{2} v$
 \item $v \leftarrow \exp \pqty{-\gamma \Delta t} v - \frac{1 - \exp \pqty{-\gamma \Delta t}}{\gamma}F + \sqrt{m} \sqrt{T_\mathrm{target} \bqty{1 - \exp \pqty{-2\gamma \Delta t}}} n$
 \item $x \leftarrow x + \frac{\Delta t}{2} v$
\end{enumerate}

\subsection{Langevin thermostat GJF}
\begin{enumerate}
 \item $v \leftarrow v + \frac{\Delta t}{2m}F$
 \item $x \leftarrow x + \frac{\Delta t}{2} v$
 \item $v_\mathrm{half} \leftarrow \frac{1}{\sqrt{1 + \frac{\Delta t \gamma}{2m}}} \pqty{v + \frac{\sqrt{2 \gamma \Delta t T_\mathrm{target}}}{2m} n}$
 \item $v \leftarrow \pqty{1 - \frac{\Delta t \gamma}{2m}} \sqrt{1 + \frac{\Delta t \gamma}{2m}} v_\mathrm{half} + \frac{\sqrt{2 \gamma \Delta t T_\mathrm{target}}}{2m} n$
 \item $x \leftarrow x + \frac{\Delta t}{2} v$
 \item $v \leftarrow v + \frac{\Delta t}{2m}F$
\end{enumerate}

\end{document}